\documentclass[pre,twocolumn,showpacs,preprintnumbers,amsmath,amssymb]{revtex4}

\usepackage{graphicx}
\usepackage{dcolumn}

\begin{document}

\title{Critical density of urban traffic}

\author{Adilton Jos\'{e} da Silva}
\email{adilton.silva@gmail.com}
\author{Borko Sto\v si\' c}
\email{borko@ufpe.br}
\affiliation{
Departamento de Estat\' \i stica e Inform\' atica, 
Universidade Federal Rural de Pernambuco,\\
Rua Dom Manoel de Medeiros s/n, Dois Irm\~ aos,
52171-900 Recife-PE, Brasil
}

\date{\today}

\begin{abstract}
A modified version of the Intelligent Driver Model was used to simulate traffic in the district of Afogados, in the city of Recife, Brazil, with the objective to verify whether the complexity of the underlying street grid, with multiple lane streets, crossings, and semaphores, is capable of exhibiting the effect of critical density: appearance of a maximum in the vehicle flux versus density curve. Numerical simulations demonstrate that this effect indeed is observed on individual avenues, while the phase offset among the avenues results in damping of this effect for the region as a whole.

\end{abstract}


\maketitle

\section{Introduction}

Over the past decades various models have been proposed for simulating traffic flow
(see e.g. \cite{Maerivoet2005,Chakroborty2006,Chandler1958,hoogendoorn2001,Brackstone1999,Chowdhury2000} and references therein), 
which range on the  phenomenon treatment detail level (microscopic, mesoscopic and macroscopic models), 
are based on individual events or global temporal behavior (discrete and/or stochastic), 
implementing real time or off-line simulations.

Among these, the so called ``car-following'' models play an important role, as they capture
much of the realism of the phenomenon, with a high level of detail.
In particular, the 
``Intelligent Driver Model'' (IDM) proposed by  Treiber and collaborators \cite{treiber1999,treiber2000}
has attracted much attention, as it is capable of reproducing diverse realistic effects.
This model is based on a system of coupled non-linear differential equations, where motion of each vehicle 
depends on the position and speed of the adjacent vehicles. The solution of this system of equations
is only possible through numerical methods. 

However, none of these models has up to date shown the ability to reproduce the 
empirically observed effect of a maximum in the vehicle flux versus density curve,
which represents a fundamental phenomenon from the point of view of urban traffic planning.

In this work we implement a modified version of the IDM model, by assuming a truncated Gaussian
distribution for the ``desired" individual vehicle speed values, on top of the real street map of the suburb of Afogados
in the city of Recife, Brazil, where we also implement real traffic rules. 
It is found that the composite effect of intermittent traffic lights, multiple merging lanes,
and random ``desired routes" of individual vehicles does produce a maximum (albeit weak) in the flux versus density curve
when a single avenue is considered as part of a larger composite system, 
whereas this effect vanes off when the measurements are made on the district system as a whole.

To this end, we have developed a simulator with a ``Graphical User Interface'' (GUI), using animation to facilitate the correction of eventual problems, visualize the operation and the simultaneous interactions, and aid in a presentation which gives credibility to the model. Most importantly, this approach enables one to interactively adapt the model variables, 
which greatly aids the realization of numerical experiments.

In the following section we give a brief overview of the necessary aspects of the traffic theory
and the IDM model, in the subsequent section we describe the numerical experiment and present the results, 
and finally, we draw the conclusions.

\section{Modeling urban traffic}

\subsection{Basic concepts}

There are many different aspects of vehicle properties, driver behavior, and the surrounding traffic infrastructure that may be
taken into account when modeling urban traffic.
For example, the vehicle weight has direct impact on the ability of changing the vehicle speed \cite{Drew1968}. The movement of a vehicle $\alpha$ in a given lane of an avenue may be described by considering the vehicle size $l_{\alpha}$, its position in relation to the lane beginning $x_{\alpha}$, speed $v_{\alpha}$ and the acceleration  $a_{\alpha}$.
In addition, individual driver characteristics influence the unique form in the vehicle conduction \cite{Drew1968, Boer1999}, where the driver perception and the reaction time are the key elements for the efficient vehicle operation \cite{Drew1968}. 
Furthermore, individual driver factors in the physiologic context \cite{Groeger1998} may be taken into account.

Generally speaking, urban traffic represents a highly complex non-linear phenomenon, 
directly related with the vehicle and driver properties, the interactions between the vehicles, 
the properties and configuration of the traffic infrastructure (streets, lanes, traffic lights etc.)
and the quantity of vehicles that enter and exit the observed region per unit time \cite{manual2000}.

The traffic flow $q$ is defined 
\cite{Maerivoet2005, manual2000, Ashton1966}
as the number of vehicles N which pass through a detector during a given time interval $\Delta t$ 
\begin{equation}
q = \frac{N}{\Delta }\ ,
\label{eq:fluxo}
\end{equation}
where in the case of multiple lanes, the composite vehicle flow on $L$ lanes is simply given by 
\begin{equation}
q = \frac{1}{\Delta t}\sum^{L}_{i=1}N_{i}\ .
\label{eq:fluxonf}
\end{equation}
Depending on the vehicle density and the street capacity, 
traffic flow may be classified as belonging to one of the three basic regimes: free flow, synchronized traffic flow (when the street capacity is approached),
and ``wide moving jam'' (when the street capacity is exhausted) \cite{Maerivoet2005, helbing2002}.

\begin{figure}
       \includegraphics[scale=0.30]{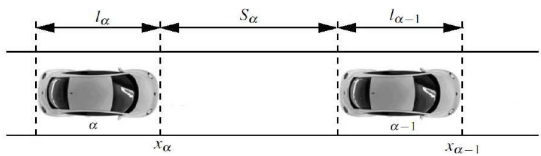}
       \caption{\label{fig:veiculos}
       Consecutive vehicles in a lane.}
\end{figure}

Vehicles in a given lane are ordered due to the impossibility of overtaking within the lane \cite{Maerivoet2005}, the first in will be the first out, unless lane switching is permitted. The distance $S_{\alpha}$ between the vehicle $\alpha$ and the vehicle $\alpha - 1$ immediately ahead, is given by the distance from the back bumper of the vehicle in front to the front bumper of the following vehicle, represented by 
\begin{equation}
S_{\alpha}  = x_{\alpha-1} - x_{\alpha} - l_{\alpha-1}\ ,
\label{eq:distancia}
\end{equation}
where $x_{\alpha}$ represents the vehicle $\alpha$ relative position within the avenue, and $l_{\alpha}$ is the vehicle length,
as presented schematically in Fig.~\ref{fig:veiculos}.
Taking the instantaneous speed of all the vehicles in the region at a given instant,
the average speed $v_{s}$ is calculated \cite{Maerivoet2005} 
as the arithmetic mean of these values
\begin{equation}
v_{s}  = \frac{1}{N} \sum^{N}_{\alpha=1}v_{\alpha}\ .
\label{eq:velmedia}
\end{equation}
Finally, the vehicle density $\rho$ is defined \cite{Maerivoet2005} as the fraction of the space occupied by all the vehicles						
\begin{equation}
\rho = \frac{1}{K} \sum^{N}_{\alpha=1}l_{\alpha}\ ,
\label{eq:ocupacao}
\end{equation}
where $K$ is the lane length.

\begin{figure}[htb]
\centering  
\includegraphics[scale=0.30]{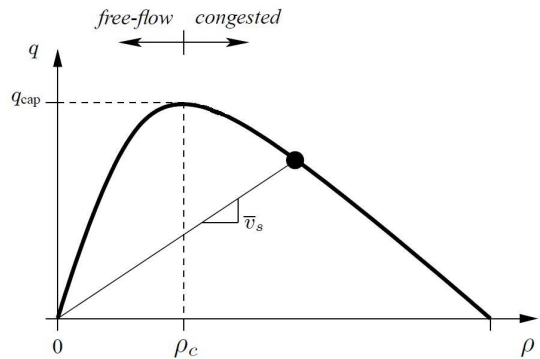}
\caption{Fundamental diagram relating the density $\rho$ to the flow $q$. The flow capacity $q_{cap}$ is reached at the critical density $\rho_{c}$. The average vehicle speed $v_{s}$ is found as the inclination of the straight line that passes through the origin.}
\label{fig:relacao}
\end{figure}

Empirical studies of traffic flow indicate a deterministic relationship between density $\rho$, flow $q$ and the average vehicle speed $v_s$ \cite{Drew1968, Maerivoet2005, manual2000, Ashton1966, Guan2008}. This relation is fundamental to the traffic flow theory, expressed by the equation $q=\rho v_{s}$ and graphically represented by the Fig.~\ref{fig:relacao}, where knowledge of two of these three variables permits determining the third. 
However, the functional relationship $q(\rho)$ remains unknown.

It is seen from Fig.~\ref{fig:relacao} that in the free traffic region, at low density values, the flow increases in a linear fashion with the  increase of density. The congested traffic begins when critical density is reached, which defines the flow capacity. From this point on, the flow begins to decrease with the density increase, corresponding to the regime of congested traffic.

At present there is no consensus in the literature as to the functional form of traffic flow versus density curve. Several approaches have been proposed up to date which end up extracting different functional forms from the same data sets. Also, there seems to be no general consent up to date as to the relevance of individual components of various existing models for the description of diverse effects of interest for the understanding of the urban vehicular traffic phenomenon.

\subsection{Intelligent Driver Model}

The ``Intelligent Driver Model'' (IDM) \cite{treiber1999,treiber2000} is governed by a system of non-linear
differential equations for the individual vehicles indexed by $\alpha$, given by
\begin{equation}
\label{eq:aceleracao}
\frac{\partial v_{\alpha}}{\partial t} = a \left[ 1 - \left(  \frac{v_{\alpha}}{v_{0}} \right) ^{\delta} - \left( \frac{S^{*}(v_{\alpha}, \Delta v_{\alpha}) } {S_{\alpha}} \right)^{2} \right],
\end{equation}
where $a$ is maximum (or intrinsic) acceleration, $\delta$ is the parameter controlling the rate at which the desired speed is approached,
 $v_{0}$ is the ``desired" vehicle speed, $S_{\alpha}$ is the distance to the preceding vehicle $\alpha-1$
\begin{equation}
\label{eq:follow}
S^{*}(v_{\alpha}, \Delta v_{\alpha}) = S_{0} + max \left(Tv_{\alpha} + \frac{v_{\alpha} \Delta v_{\alpha}}{2\sqrt{a b}} , 0\right) ,
\end{equation}
is the ``safety distance" between the vehicles $\alpha$ and $\alpha-1$, $S_{0}$ is the minimum security distance, $T$ is the breaking reaction time \cite{Leutzbach1988}, $\Delta v_{\alpha}$ is the difference between the velocities of the two vehicles, and $b$ is the deceleration parameter governing the influence of the velocity difference on the safety distance.

Beside an abundance of parameters (perhaps, over-abundance for the current purpose), this model contains intuitively attractive components for the description of vehicular behavior, and has been shown to display rather
realistic behavior in diverse situations (see \cite{treiber1999,treiber2000} and citing references). 
The solution
of the model does not present physically unacceptable situations (such as negative velocities, vehicle overlapping etc.), and stable solutions
are attainable already with the Euler's method, while application of the Runge-Kutta method yields accurate results for describing effects such as e.g. traffic density waves propagation.

On the other hand, this model yields realistic effects only in the presence of obstacles or conflicting situations such as confluent lanes,
whereas in the case of free lane traffic it demonstrates trivial behavior (all vehicles transit in parallel with the same speed), without limit for the street capacity. It turns out that using a truncated Gaussian distribution \cite{Walck2007, boxmuller1958} for maximum desired speeds of individual vehicles, is sufficient to create a traffic congestion in an avenue without obstacles  \cite{Santos2008}, where street capacity becomes apparent through saturation of the flow curve.
However, the maximum of the flow curve at the critical density was not observed with this generalization of the original IDM model \cite{Santos2008}.

In order to investigate whether this phenomenon may be brought about by the additional confusion stemming from the complexity of the environment (multiple lanes, vehicle types, traffic rules etc.), in what follows we implement a numerical simulation implementing such a situation.

\section{Numerical experiment}

The area of study is located in the Recife Metropolitan Region (RMR), where a network of urban highways connects municipal districts and suburbs. According to Pernambuco State Traffic Department (DETRAN) data, in the year of $1990$ the RMR was traversed by some $250000$ vehicles, which has grown to over $700000$ in  $2007$, where over $400000$ of these are found in the city of Recife alone. This violent vehicular density increase 
has been causing heavy traffic jams in several parts of the city over the past years, augmenting the attention of the authorities and researchers.
The study region itself represents one of the two principal transport corridors between the south metropolitan region and Recife downtown.

As already mentioned, the IDM model by itself does not yield results demonstrating critical behavior on the flux versus density scatter plot,
and the generalization of IDM which considers a distribution of desired velocities leads only to saturation of the flux versus density curve,
with no sign of diminishing flux when the density is further increased. In order to verify whether the complexity of an urban traffic system is able
to reproduce such an effect, in the current numerical experiment we have implemented various components encountered in the real life situation. 
In particular, we consider the following basic elements:
\begin{itemize}
	\item The avenues are considered  in terms of length, number of lanes, crossings, and traffic lights;
  \item	Vehicles of different size and maximum acceleration are considered (trucks, busses and cars), with individual maximum and minimum ``desired" speed. Upon entry, each vehicle is (randomly) assigned a ``desired" itinerary.
  \item	Traffic rules are implemented in each avenue and lane, maneuvering is implemented in accord with the existing signalization, considering the possibility of changing lanes and overtaking slower vehicles, while avoiding collisions.
\end{itemize}
\begin{figure}[htb]
       \centering  
       \includegraphics[width=3.4in]{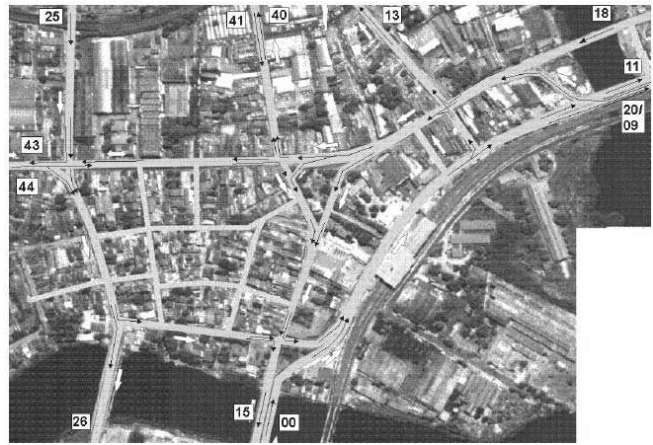}
       \caption{Street map of the Afogados suburb with indicated traffic flow directions.}
       \label{fig:afogados}
\end{figure}
We present in Fig.~\ref{fig:afogados} the street map of the Afogados suburb with six entries and seven exits, where the directions of the traffic flow are indicated.
All the avenue crossings were visited and examined physically, in order to take into account all the possible maneuvering options of a vehicle
in a given lane of a given avenue, which attempts to follow a given route, while honoring the existing traffic rules.

The traffic lights were modeled as elements that belong to the avenues, such that the vehicles detect the traffic light presence, and behave in accord with the stop or go state (in fact, this was implemented by inserting a virtual still vehicle at the crossroad when the traffic light is red, such that approaching vehicles sense its presence, and break according to the IDM rules).

\begin{figure}[htb]
       \centering  
       \includegraphics[width=3.4in]{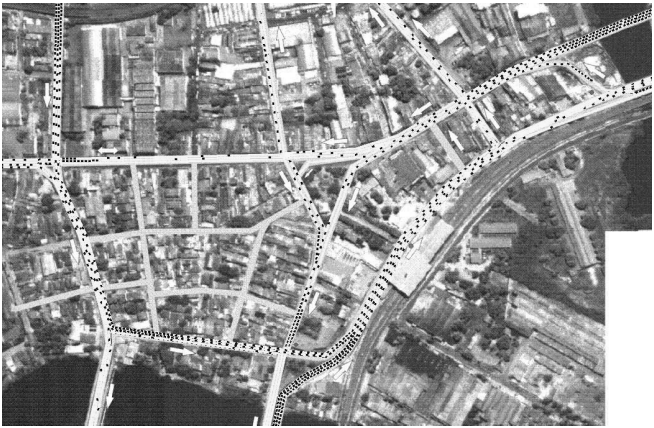}
       \caption{A snapshot of the graphic interface of the simulator during the simulation.}
       \label{fig:simulacao}
\end{figure}

As already mentioned, the vehicles entering the region are attributed (random) desired routes. In the attempt to follow this route, a vehicle chooses permitted maneuvers, attempting to choose a lane that optimizes access to a determined crossroad, and realizes lanes changing and overtaking in the attempt to reach the destination. To this end we implement a origin $\rightarrow$ destination route table taking into account all the possible  combinations of the entry and exit avenues that comply with the traffic rules.
While attempting to follow the desired route, the vehicle may also change lanes due to advantageous conditions (e.g. when directly in front of it there is a slower vehicle, and the adjacent lane is empty), while honoring traffic rules. 
This basic setup creates a rather realistic behavior of individual vehicles, and it seems to capture the collective behavior
of the vehicle fleet, which is observed on the graphical interface, as shown in Fig.~\ref{fig:simulacao} .

In order to achieve a real time simulation on the current scale that permits realistic traffic behavior, we have opted for the Euler algorithm
with a $0.05s$ time increment, and a full scene refresh at the rate of two frames per second. While a higher order Runge-Kutta method yields higher
precision results, the computational demand on this simulation scale takes it out of the real time observation range on the current hardware, and we have verified that the overall behavior of the traffic system is not affected by the precision gain.

For the model parameters we adopt the values $a=1.5 m/s^2$, $b=2.0 m/s^2$, $T=1.2 s$, $S_{0}=2.0 m$, and $\delta =4$, while the values of the desired maximum speed for each vehicle are drawn from a normal distribution truncated at $V_{min}$ and $V_{max}$,
depending on the vehicle type, and in accordance with the local traffic rules,
with parameters given in Tab.~\ref{tab:velocidade}.
\begin{table}[h]
	\centering
		\caption{
		Parameters 
		of the normal distribution truncated at $V_{min}$ and $V_{max}$, 
		used to draw the desired maximum speed for different vehicle types 
		(the values are given in $km/h$).
		}
		\begin{tabular}{l c c c c c}
		\hline
		Vehicle type & $V_{max}$ & $V_{min}$ & $\mu$ & $\sigma$ \\
		\hline
		cars              & 80 & 40 & 60  & 20 \\
		buses  & 50 & 30 & 40  & 5  \\
		trucks    & 40 & 30 & 35  & 2.5  \\
		\hline
		\end{tabular}
	\label{tab:velocidade}
\end{table}

The simulation is performed by inserting vehicles into all the incoming avenues (and lanes) at the edge of the simulation zone. 
The insertion rate is gradually increased from $0.125$ cars per second per lane ($8 s$ delay between incoming vehicles), to continuous influx (no delay between inserting the vehicles), with successive decrements of $0.1 s$ for the delay between incoming vehicles. At each rate (altogether eighty values), ample equilibration time of $1800 s$ ($30 min$) is used to attain equilibrium, after which the vehicle count and individual speeds are recorded in each avenue segment bounded by street crossings and confluence points. This process was repeated $100$ times in order to establish the average values.

\section{Results}

\begin{figure}
\centering
{  
  \includegraphics[width=0.4\textwidth]{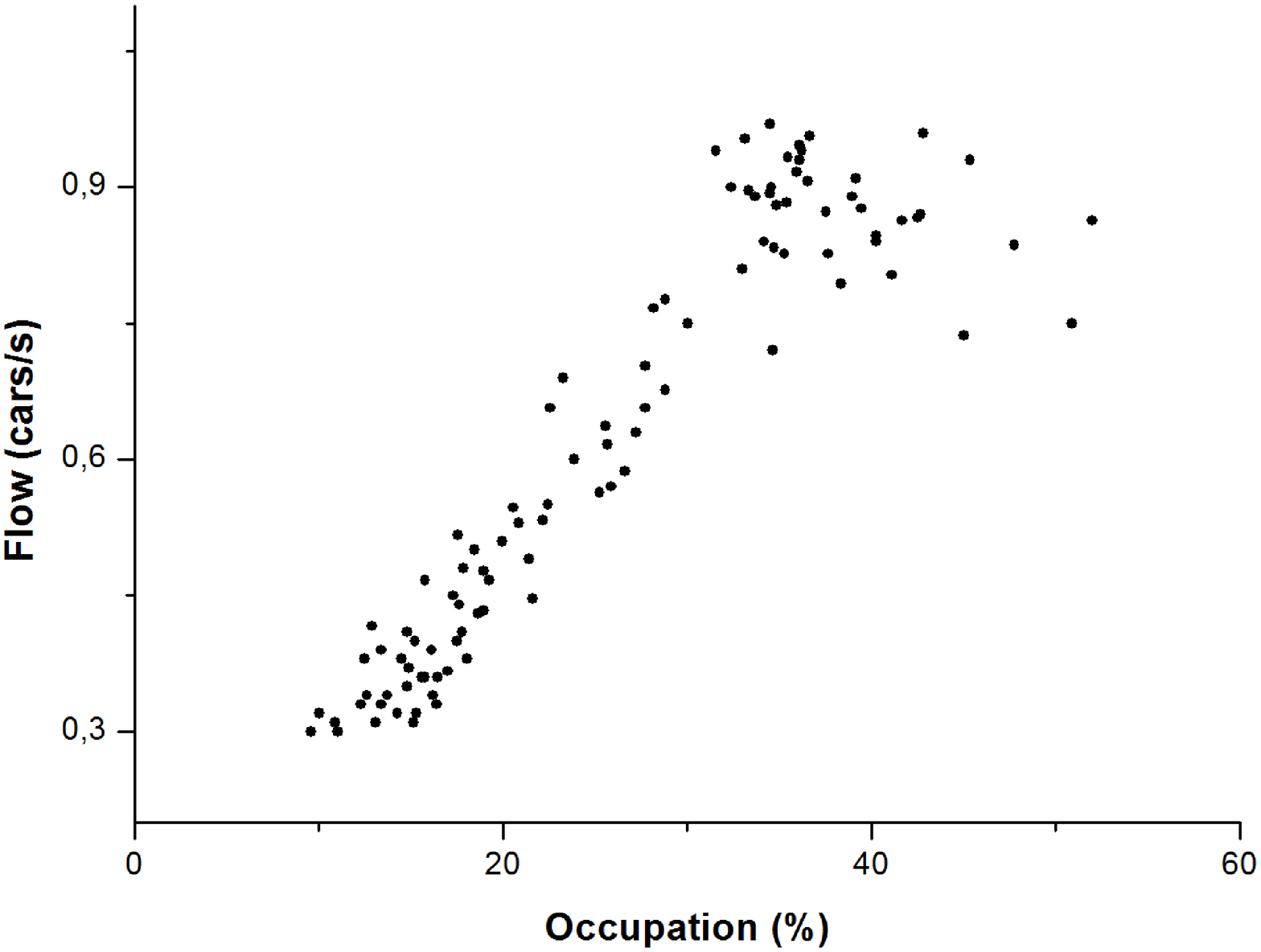}}    
  \includegraphics[width=0.4\textwidth]{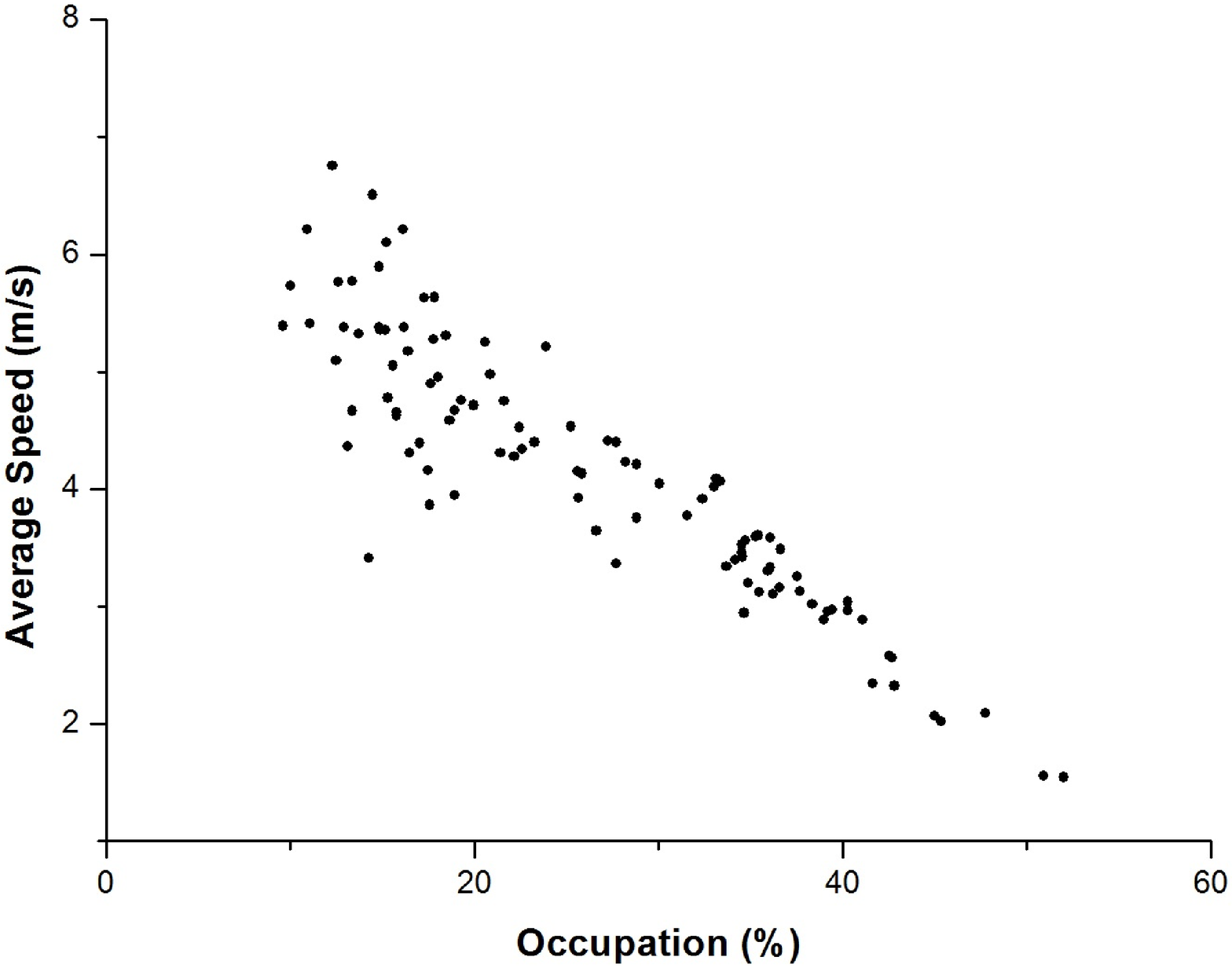}
  \caption{Simulation results for the avenue $00$.}
  \label{fig:results00}
\end{figure}

Results of the simulation corresponding to the avenue labeled ``$00$" in Fig.~\ref{fig:afogados} are presented in Fig.~\ref{fig:results00},
where it is observed that the flux versus occupation density scatter plot does display signs of a maximum (and thus existence of critical density),
while the average speed versus occupation demonstrates large fluctuations in the low density regime, which diminish with density increase.

\begin{figure}[h]
       \centering  
       \includegraphics[scale=0.23]{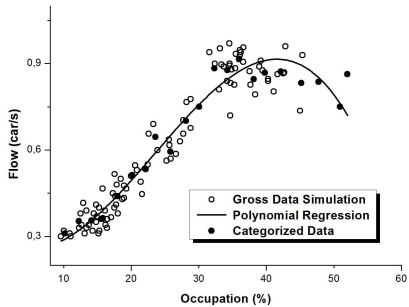}
       \caption{Polynomial regression of the third order for simulation data, avenue $00$.}
       \label{fig:regressao}
\end{figure}

In order to alleviate the effect of large fluctuations and pinpoint the critical occupation density, in Fig.~\ref{fig:regressao} we display the same data as in Fig.~\ref{fig:results00}, together with the results of averaging the data in $2\%$ bins, and polynomial regression which indicates a critical occupation of $41.63\%$, with a maximum flow of $0.91998\; cars/s$ (or $3312\; cars/h$).

\begin{figure}[h]
  \centering
  \includegraphics[width=0.45\textwidth]{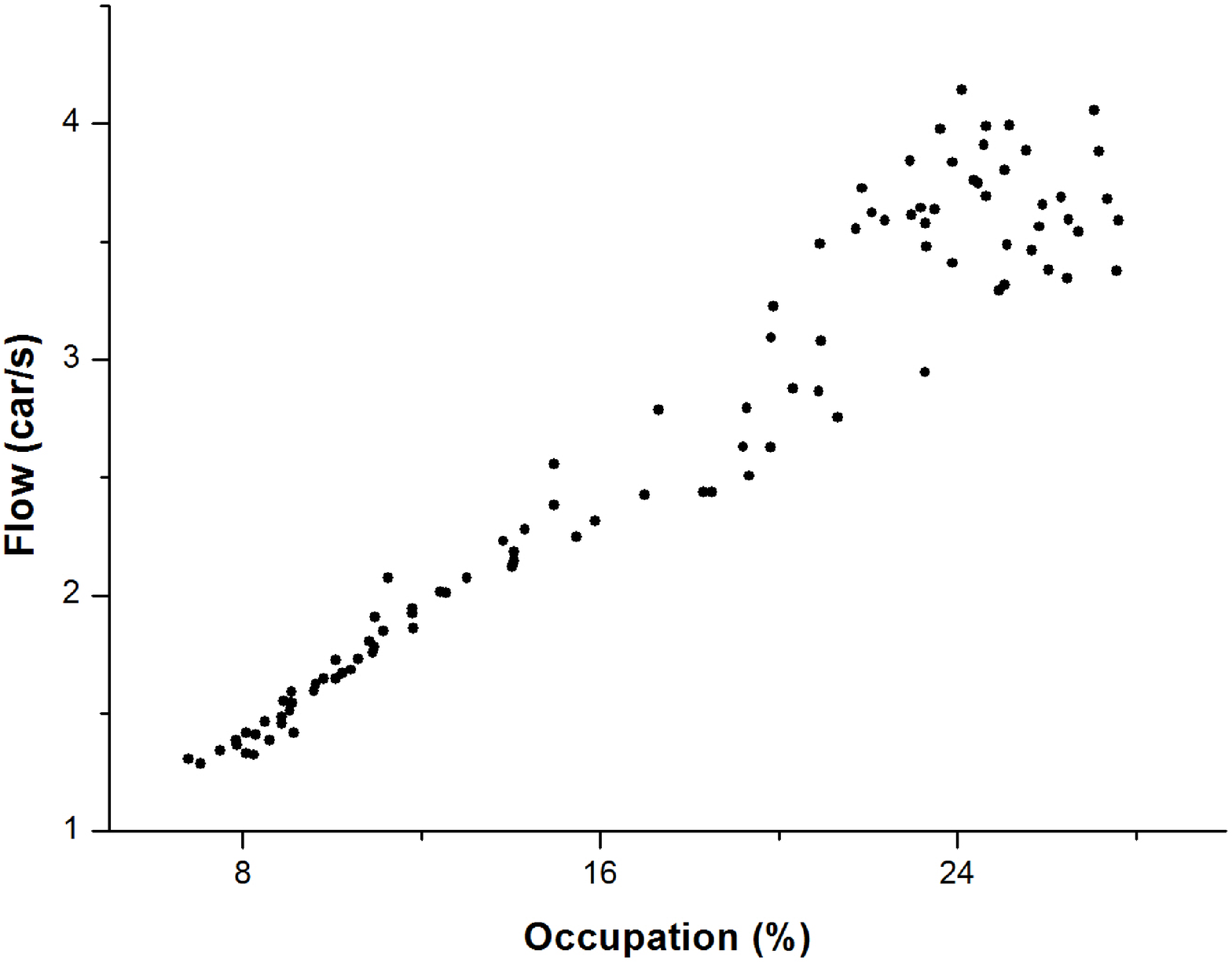}\label{fig:ocupacaogeral}
  \includegraphics[width=0.45\textwidth]{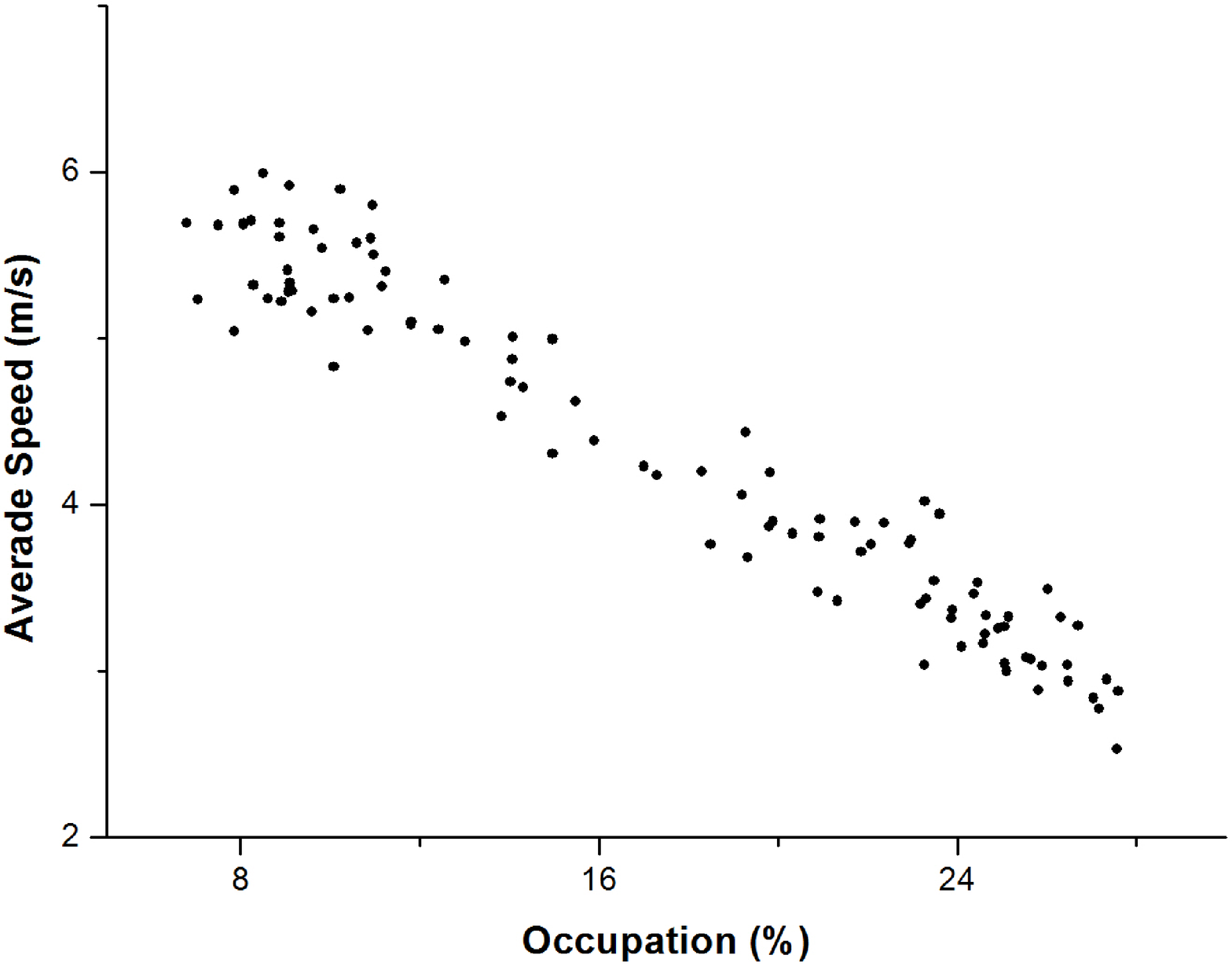}\label{fig:velocgeral}
  \caption{Collected gross data for the entire region.}
  \label{fig:geral}
\end{figure}

Results integrated over the entire simulation region are displayed in Fig.~\ref{fig:geral},
which does not reveal a maximum in the flux versus occupation density scatter plot. Also, variation of
the average velocity is pronounced over the whole observed range of density occupation values.
This findings may be attributed to the fact (observed on the graphical interface during the simulation)
that the entire region never becomes entirely congested, rather, different regions take turns in
demonstrating free flow, and traffic jam states.

\section{Conclusions}

In this work we implement a modified version  of the Intelligent Driver Model on top of a real street map of a suburb in Recife, Brazil,
in order to verify whether the additional complexity brought about by the synchronized collective behavior of vehicles that attempt to follow
individual routes, is capable of producing the critical occupation density effect with a maximum of the flux versus density curve. This
behavior is empirically observed, and represents a fundamental effect from the point of view of urban traffic planning, but has not been reproduced
up to date by existing mathematical and numerical models.

The modification of the IDM model implemented in this work consists of several ingredients. In particular, we consider three types of vehicles
with different characteristics (length, maximum acceleration, and individual desired speed), multiple lanes with traffic rules (merging traffic, semaphores etc.), and predefined individual routes for every vehicle.

It turns out that different parts of the observed region go intermittently through phases of free flow and congested traffic behavior. If only a single avenue is observed, the composite effect of interaction with the neighboring regions is reflected in behavior that may be identified as critical 
density (albeit weak in the current simulation). However, the phase offset between the regions (while a given region is congested, a neighboring region displays free flow, and vice versa) leads to canceling out of individual congestion effects, and the maximum in the flux versus density curve is not observed. Also, the average velocity as a function of occupation displays heteroscedastic behavior, with large fluctuations at low densities that diminish as the density is increased, whereas large average velocity fluctuations are observed for the entire occupation density range for the suburb as a whole.

We may conclude that the critical density phenomenon should be regarded as a local effect, brought about by the interplay of a given region with
the neighboring regions, as an exit point for one region represents an entry point for another. To the best of our knowledge, the current work represents the first report in the literature regarding a mathematical/numerical model capable of exhibiting this effect.

Further studies should be made as to the contribution of each of the implemented model components to the observed critical density effect, and whether it may be enhanced by inclusion of some others. 
It should also be investigated whether other microscopic models are capable of demonstrating this phenomenon in a similar setup.


\section{Acknowledgements}
The authors acknowledge financial support of Brazilian federal agencies CNPq (project CASADINHO-620113/2008) and
CAPES (projects PROCAD-1396/2007 and PROCAD-2273/2008).



\end{document}